\documentclass{aa}
\usepackage{graphicx,natbib}
\bibpunct{(}{)}{;}{a}{}{,}

\def\grs{GRS\,1734$-$292}
\def\egret{3EG\,J1736$-$2908}

\newcommand{\lesssim}{\mathrel{\hbox{\rlap{\hbox{\lower4pt\hbox{$\sim$}}}\hbox{$
<$}}}}
\newcommand{\gtrsim}{\mathrel{\hbox{\rlap{\hbox{\lower4pt\hbox{$\sim$}}}\hbox{$>
$}}}}

\begin{document}

\title{Broadband X-ray spectrum of GRS\,1734--292, a luminous Seyfert~1
galaxy behind the Galactic Center}

\author{S.Yu. Sazonov \inst{1,2}, M.G. Revnivtsev \inst{1,2}, A.A. Lutovinov
\inst{2}, R.A. Sunyaev\inst{1,2} and S.A. Grebenev  \inst{2}}

\offprints{sazonov@mpa-garching.mpg.de}

\institute{Max-Planck-Institut f\"ur Astrophysik,
           Karl-Schwarzschild-Str. 1, D-85740 Garching bei M\"unchen,
           Germany
     \and   
           Space Research Institute, Russian Academy of Sciences,
           Profsoyuznaya 84/32, 117997 Moscow, Russia
}
\date{}

\authorrunning{Sazonov et al.}
\titlerunning{Broadband X-ray spectrum of GRS\,1734--292}
 
\abstract{Based on a deep survey of the Galactic Center region
performed with the INTEGRAL observatory, we measured for the first
time the hard X-ray (20--200\,keV) spectrum of the Seyfert 1
galaxy \grs\ located in the direction of the Galactic Center. We extended the
spectrum to lower energies using archival GRANAT and ASCA
observations. The broadband X-ray spectrum is similar to those of
other nearby luminous AGNs, having a power law shape without cutoff up
to at least 100\,keV. 
}

\maketitle

\section{Introduction}

The hard X-ray source \grs, located $1.8^\circ$ from the Galactic
Center (GC), was discovered in 1990 by the ART-P telescope 
aboard the GRANAT satellite \citep{pgs92,pgs94}. The measured power-law
spectrum with a photon index $\Gamma\sim 2$ and inferred X-ray luminosity 
($\sim 10^{36}$\,erg\,s$^{-1}$) assuming a GC distance were consistent
with the source being a Galactic X-ray binary. However, subsequent optical
spectroscopic observations surprisingly revealed \citep{mmc+98} very
strong, broad emission lines demonstrating that \grs\ is the nucleus
of a Seyfert 1 galaxy at a redshift $z=0.0214$. The galaxy itself has so
far escaped observation because of $\approx 6$ magnitudes of visual
absorption along the line of sight through the Galactic plane. 

With its X-ray luminosity approaching $10^{44}$\,erg\,s$^{-1}$
(2--10\,keV), \grs\ is one of the $\sim 5$ most luminous AGNs within
100\,Mpc of us \citep[e.g.][]{pmb+82,sr04}, which makes it a very
interesting object for investigation.  Additional interest in \grs\ is
connected with the 
fact \citep{dcc+04} that its position falls into the error box ($\sim
0.5^\circ$ radius) of the gamma-ray source \egret\ discovered by
CGRO/EGRET \citep{hbb+99}. 

In the past, spectroscopy of \grs\ at energies above
20\,keV could not be performed with collimator instruments such as
RXTE/PCA, RXTE/HEXTE and BeppoSAX/PDS because of the high number
density of bright sources in the GC region. The only instrument
possessing the necessary angular resolution was the coded mask SIGMA
telescope aboard GRANAT. However, \grs\ persistently remained below the
sensitivity threshold of SIGMA and was marginally detected only during
an outburst  on September 15--17, 1992 \citep{cgc+92}. 

\grs\ has recently been detected with high significance by the
hard X-ray imager IBIS aboard the INTEGRAL satellite
\citep{rsv+04,bms+04}. Here we use INTEGRAL observations to obtain for
the first time a high-quality X-ray spectrum of \grs\ above
20\,keV and extend the spectrum to lower energies using data from
previous missions.  

\section{Observations and data analysis}

The main instruments of INTEGRAL are the hard X-ray coded mask
telescope IBIS and spectrometer SPI \citep{wcd+03}. Unfortunately, the
limited angular resolution ($\sim 2.5^\circ$) does not allow us to use
SPI for studying \grs, because 
there are several bright hard X-ray sources, including
1E\,1740.4$-$2942, located within a few degrees of \grs. We therefore
employ the IBIS telescope (specifically the ISGRI detector) whose high
angular resolution ($\sim 10^\prime$) prevents source confusion. 

A $30^\circ\times 30^\circ$ area centered on the GC was
extensively observed by INTEGRAL in August--September 2003, these
observations making up an ultra-deep survey with a total exposure of
$\sim 2$\,Ms. A total of 60 point sources were detected by
IBIS/ISGRI, including \grs\ \citep{rsv+04}.   

Here we have analyzed the data of the GC survey
following the methods described by \cite{rsv+04}. Specifically, 
source spectra are obtained by building images in a set of energy
intervals followed by normalizing the resulting source fluxes 
to the corresponding fluxes of the Crab for a similar position
in the field of view. Our analysis of an extensive set of Crab
calibration observations has shown that the source absolute flux can be 
recovered with an accuracy of 10\% and the systematic uncertainty of
relative flux measurement in different energy channels is less than 5\%.

In the 2--10\,keV band we used a 6-ks observation performed on
March 12, 1999 with the ASCA/GIS telescope \citep{skm+02}. The data
reduction was done using the LHEASOFT/FTOOLS 5.2
package. In the 3--20\,keV range, partially overlapping with the ASCA
and INTEGRAL bands, we used the published spectrum based on GRANAT/ART-P
observations of \grs\ in September--October 1990 \citep{pgs94}. 

\section{Results}

\begin{figure}
\includegraphics[bb=15 180 580
720,width=\columnwidth]{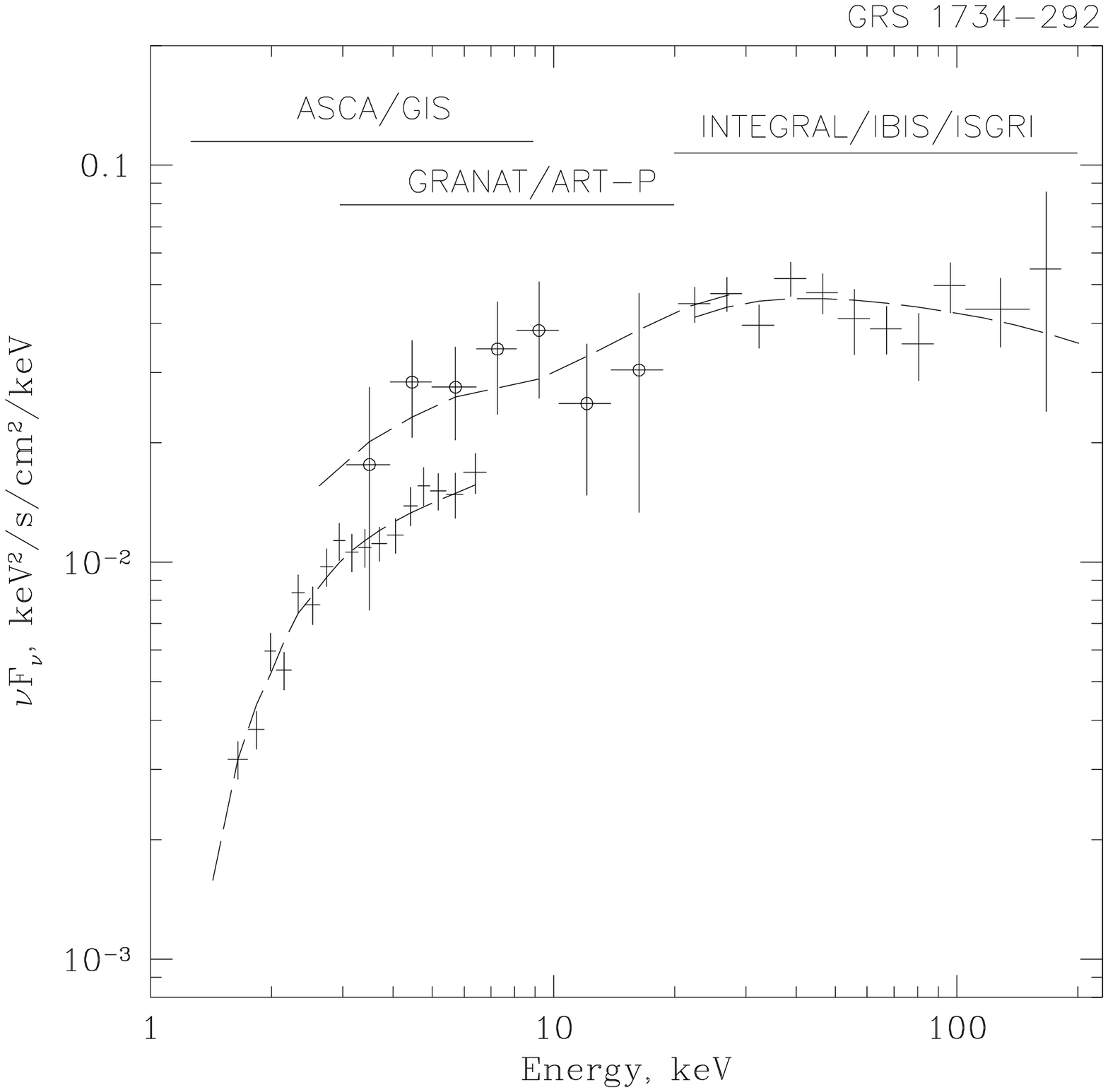} 
\caption{Composite X-ray spectrum of \grs\ obtained from
nonsimultaneous INTEGRAL, GRANAT and ASCA observations. The dashed
lines indicate the best-fit (absorbed power-law+reflection) model
defined by parameters given in Table~\ref{fits} (the last set), scaled 
to match the different data sets.
}
\label{spectrum}
\end{figure}

The composite X-ray (2--200\,keV) spectrum of \grs\ is shown in
Fig.~\ref{spectrum}. We modeled the spectrum using the XSPEC package and 
the results are presented in Table~\ref{fits}. 

\begin{table}
\caption{Results of spectral analysis}
\begin{tabular}{l|l}
\hline
\hline
Power law (PL) with cutoff & IBIS\\
\hline
$\Gamma$ & $2.1\pm0.1$\\
$E_{\rm cut}$, keV & $> 110$ ($2\sigma$)\\
Flux (18--200\,keV)$^{\rm a}$ & $1.65\pm0.2$\\
Flux (2--10\,keV)$^{\rm b}$ & 1.3\\
$\chi^2/{\rm d.o.f}$ & 7.6/10 \\
\hline
Absorbed PL & GIS\\
\hline
$N_{\rm H}$, $10^{22}$\,cm$^{-2}$ & $1.5\pm0.2$\\
$\Gamma$ & $1.48\pm0.15$\\
Flux (2--10\,keV) & 0.35\\
$\chi^2/{\rm d.o.f}$ & 17.0/15 \\
\hline
Absorbed PL with cutoff & IBIS+GIS+ART-P\\
\hline
$N_{\rm H}$, $10^{22}$\,cm$^{-2}$ & $2.0\pm0.3$\\
$\Gamma$ & $1.7\pm0.2$\\
$E_{\rm cut}$ & $156^{+\infty}_{-55}$\\
Flux (2--200\,keV, IBIS)$^{\rm c}$ & 2.1\\
Flux (2--200\,keV, GIS)$^{\rm c}$ & 1.3\\
Flux (2--200\,keV, ART-P)$^{\rm c}$ & 2.1\\
$\chi^2/{\rm d.o.f}$ & 34.1/34\\
\hline
Absorbed PL with cutoff+reflection & IBIS+GIS+ART-P\\
\hline
$N_{\rm H}$, $10^{22}$\,cm$^{-2}$ & $2.2\pm0.2$\\
$\Gamma$ & $1.9\pm0.2$\\
$E_{\rm cut}$ & $>120$ ($2\sigma$)\\
$R\equiv\Omega/2\pi$ & 1.0 (fixed)\\
Flux (2--200\,keV, IBIS)$^{\rm c}$ & 2.2\\
Flux (2--200\,keV, GIS)$^{\rm c}$ & 1.8\\
Flux (2--200\,keV, ART-P)$^{\rm c}$ & 2.8\\
$\chi^2/{\rm d.o.f}$ & 34.4/34\\
\hline
\end{tabular}
\begin{list}{}
\item $^{\rm a}$ All quoted fluxes are observed (uncorrected for
absorption) ones and are given in units of
$10^{-10}$\,erg\,s$^{-1}$\,cm$^{-2}$.   
\item $^{\rm b}$ Flux in the 2--10\,keV band computed from the model
assuming interstellar absorption with $N_{\rm H}=1.5\times
10^{22}$\,cm$^{-2}$.
\item $^{\rm c}$ Model flux in the 2--200\,keV band scaled to match
the IBIS, GIS or ART-P data. 
\end{list}
\label{fits}
\end{table}

The spectrum measured at 18--200\,keV with INTEGRAL/IBIS is well fit
by a power law with a photon index $\Gamma=2.1\pm 0.1$ (hereafter all
quoted uncertainties are $1\sigma$). The ASCA/GIS
observation at 2--10\,keV indicates a harder power law, 
$\Gamma=1.48\pm 0.15$, and requires the inclusion of neutral 
absorption with a column density $N_{\rm H}=(1.5\pm 0.2)\times
10^{22}$\,cm$^{-2}$.  This column does not exceed significantly
the Galactic interstellar absorption in the direction of \grs\
($N_{\rm H}\approx 1.0\times 10^{22}$\,cm$^{-2}$) as estimated by
\cite{mmc+98}, taking into account the uncertainty in the latter value. Our
results for the ASCA observation are in good agreement with those
previously reported by \cite{skm+02}.  

An absorbed power law model with $\Gamma=1.7\pm 0.2$ modified by an
exponential cutoff with $E_{\rm cut}\gtrsim 160$\,keV provides a good fit
to the ASCA, ART-P and IBIS data combined if allowance is made
for the different flux levels in these observations. Such type of 
spectra is known to result from Comptonization of low energy
radiation in a hot plasma \cite[e.g.][]{st80}. Alternatively, the broadband
spectrum of \grs\ can  be well fit by an absorbed power law with a
Compton reflection component (pexrav model in XSPEC) whose amplitude
$R$ is poorly constrained by the data; fixing $R=1$ yields $\Gamma=1.9\pm 0.2$.

\begin{figure}
\vbox{
\includegraphics[bb=15 180 580 510, width=\columnwidth]{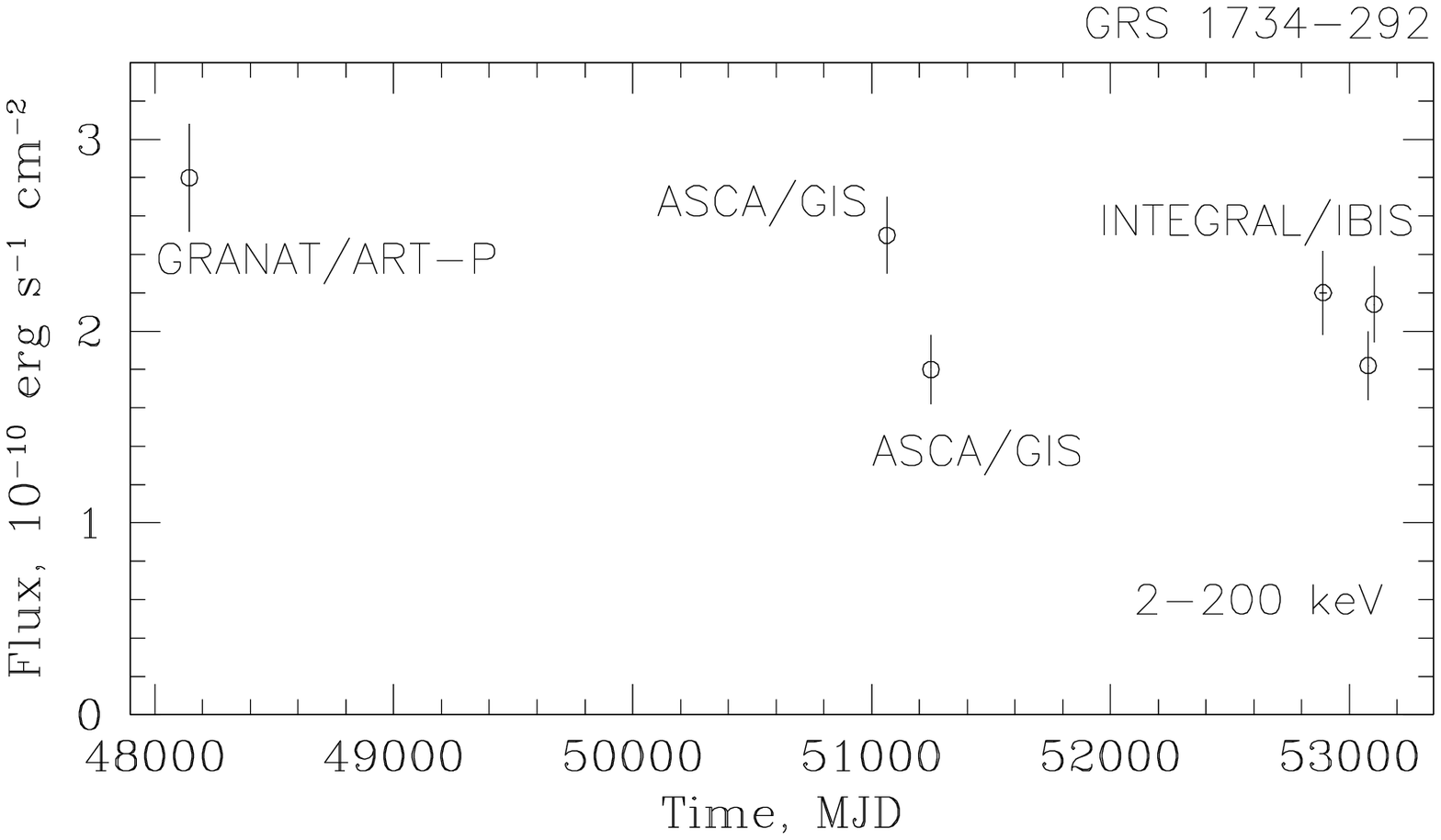}
\includegraphics[bb=15 180 580 460, width=\columnwidth]{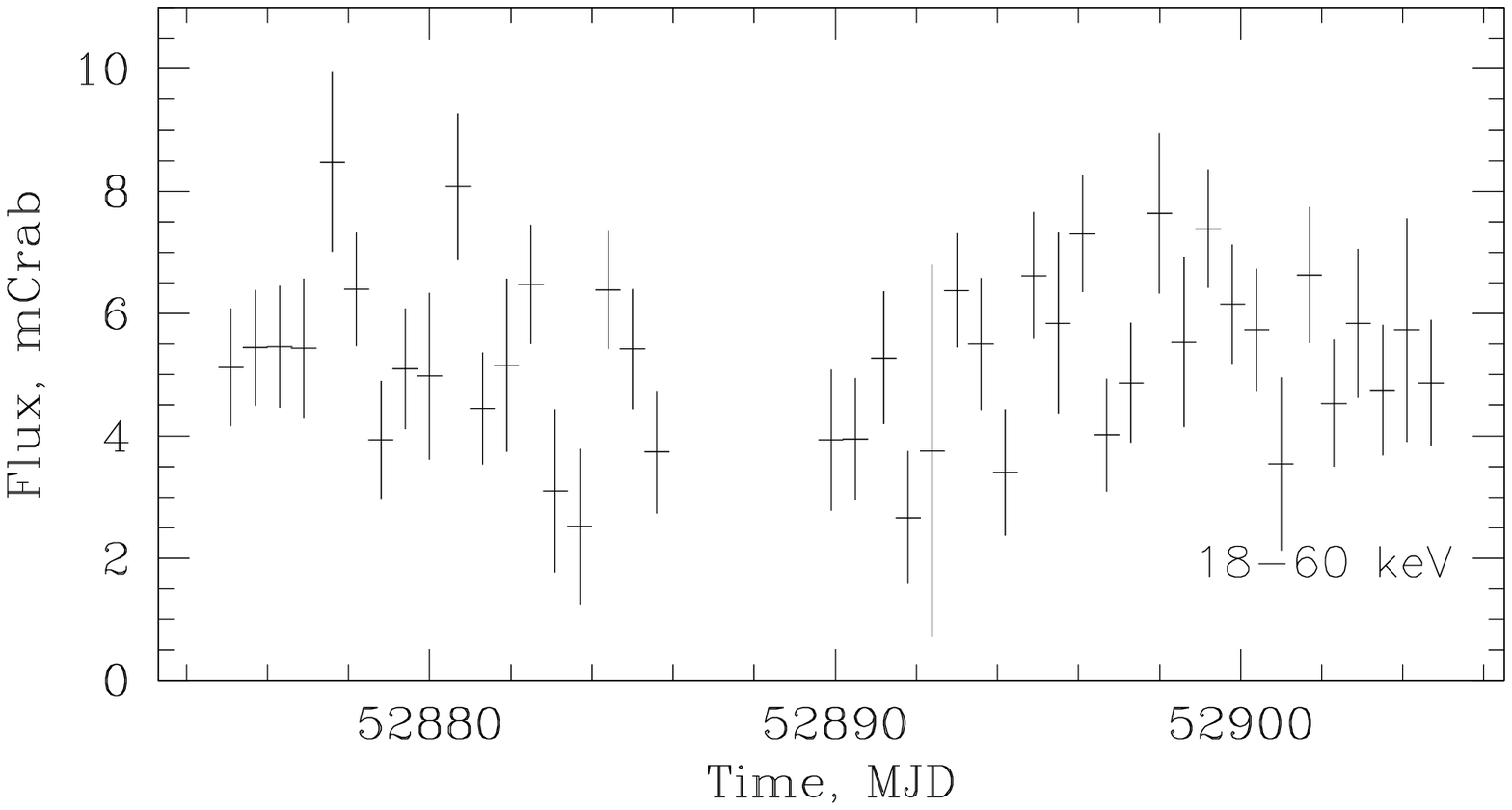}
}
\caption{{\bf Upper panel}: Time history of the observed X-ray flux from \grs\
in 1990--2004 assuming that its broadband X-ray spectrum remained
invariably as that shown in Fig.~\ref{spectrum}. {\bf Lower panel}:
Light curve of \grs\ in August--September 2003 measured with 
INTEGRAL/IBIS.
}  
\label{lcurve}
\end{figure}

The results of the above analysis are consistent with the hypothesis
that the X-ray spectrum was not significantly different in 
the ART-P, ASCA and INTEGRAL observations. Assuming a constant
spectral shape, the X-ray flux from \grs\ varied by less than a factor
of 2 from one observation to another. Figure~\ref{lcurve} summarizes
our knowledge of the X-ray light curve of the source for the period
1990--2004. This light curve includes three additional flux measurements: 
one based on an ASCA/GIS observation on September 8, 1998
and reported by \cite{skm+02}, and two based on new 
observations of the GC region with INTEGRAL/IBIS on March 12--16 and 
April 8--9, 2004 (with effective exposures of 137\,ks and 108\,ks,
respectively). The fluxes from the different instuments have been
translated to the 2--200\,keV range assuming the absorbed power
law+reflection model from Table~\ref{fits}. 

The presented long-term light curve suggests that \grs\ is fairly
stable on a time scale of years, with a typical absorption-corrected
flux of 0.7 (2.2)$\times 10^{-10}$\,erg\,s$^{-1}$\,cm$^{-2}$ in the
2--10\,keV (2--200\,keV) band. We note that \grs\ is marginally
(4\,$\sigma$) detected on a similar flux level on the cumulative  map
of the GC region obtained from Mir/Kvant/TTM observations in 
1987--1997 (M. Gilfanov,  private communication). Therefore, the typical 
intrinsic luminosity of \grs\ is $6 (20)\times
10^{43}$\,erg\,s$^{-1}$ at 2--10\,keV (2--200\,keV) (assuming
$H_0=75$\,km\,s$^{-1}$\,Mpc$^{-1}$). During the $\sim 3$-day
long outburst detected by GRANAT/SIGMA \citep{cgc+92} \grs\ was
apparently brighter by a factor of $\sim 5$. 

The 1-month series of INTEGRAL observations in 2003 allows us to study the
variability of \grs\ on time scales of days (see
Fig.~\ref{lcurve}). The source proves to be variable, with a
fractional variability amplitude of $12\pm6$\% at frequencies higher
than $6\times10^{-7}$\,Hz. These results demonstrate that \grs\ is
similar in its long-term and short-term variability properties to
other Seyferts \citep[e.g.][]{mev03}.

\section{Discussion}

The obtained X-ray spectrum of \grs\ is typical for Seyfert
galaxies. Previous hard X-ray missions including GRANAT, CGRO, RXTE
and BeppoSAX demonstrated that X-ray spectra of Seyferts can be
described by a power law of $\Gamma\approx 1.8$ modified by Compton
reflection at 10--100\,keV and an exponential cutoff at $E_{\rm
cut}\gtrsim$ 100--200\,keV
\citep{jbb+92,esm00,gzj+96,pmc+02}. Measurement of the cutoff energy
in individual objects remains a difficult task even for INTEGRAL, 
feasible only for very long exposures of the brightest
AGNs. Accumulation of statistics on the distribution of $E_{\rm cut}$
values and its dependence on luminosity (and possibly other
characteristics)  is crucial for constraining the physical parameters
of the hot plasma surrounding supermassive black holes as well 
as for a better understanding of the origin of the cosmic X-ray
background (CXB). The lower limit $E_{\rm cut}\gtrsim 100$\,keV
obtained here for \grs\ is thus valuable given the fact that this
source is one of the most luminous AGNs for which such an
estimate has been made and also because the bulk of the CXB appears to
be produced by AGNs with similar ($\sim 10^{44}$\,erg\,s$^{-1}$)
luminosities \citep{uao+03}.  

\begin{figure}
\includegraphics[bb=15 180 580
720,width=\columnwidth]{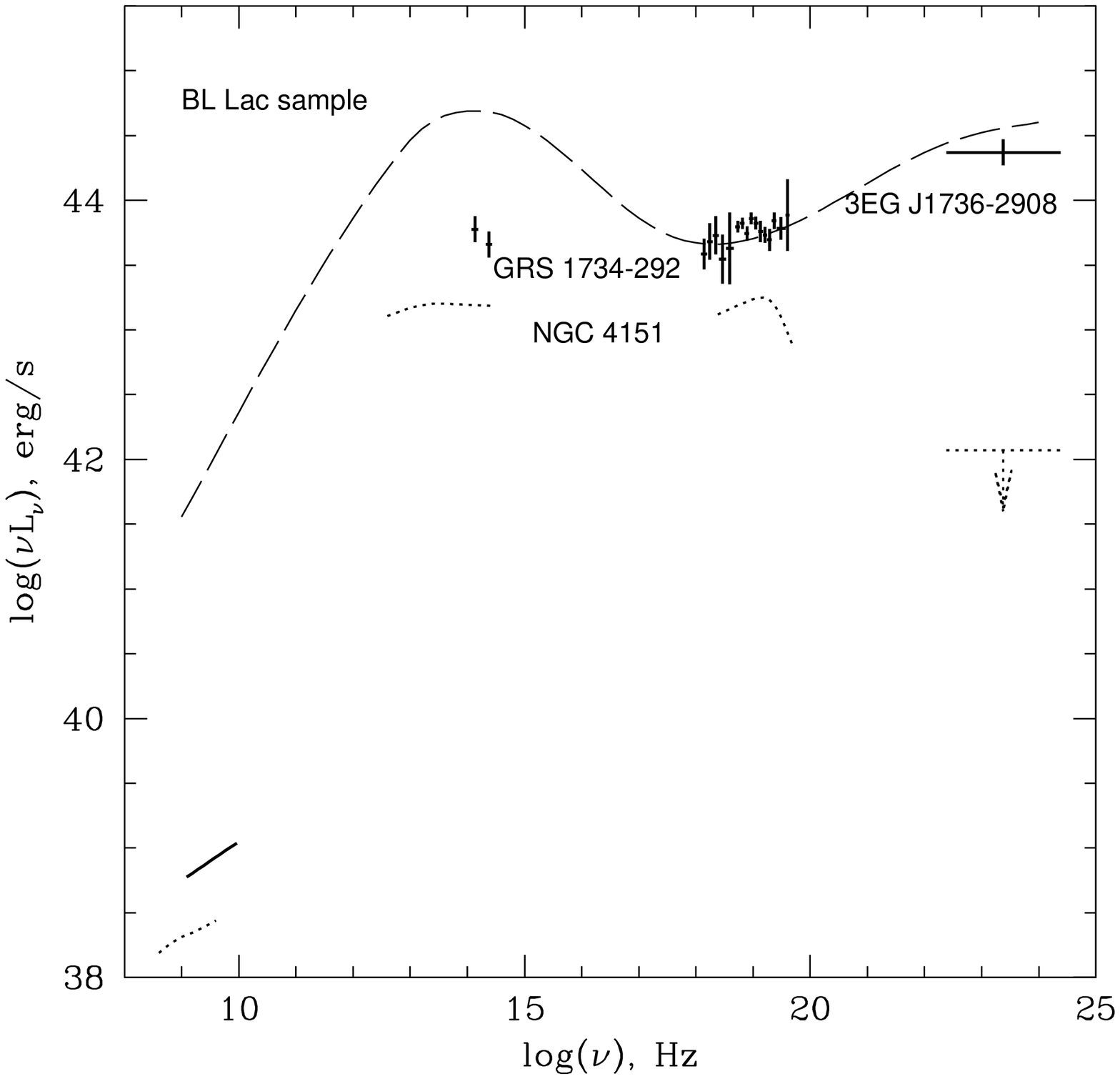}
\caption{Comparison of the broadband spectral energy distribution of
\grs\ (in solid) with those of NGC~4151 (in dotted) and of BL Lac
objects (in dashed). The data for \grs\ include the radio spectrum and near 
infrared measurements (K and H bands) adopted from \cite{mmc+98},
X-ray spectrum at 5--200\,keV from GRANAT/ART-P and INTEGRAL/IBIS and
the gamma-ray flux measured from \egret\ with CGRO/EGRET in
July--August, 1992 \citep{hbb+99}. For NGC~4151, the radio to optical
spectrum is approximated from multifrequency data taken from the NED
database, the X-ray spectrum above 10\,keV is 
adopted from \cite{fcg+95}, the upper limit on the flux above 100\,MeV
is taken from \cite{lbd+93}, and a distance of 20\,Mpc is assumed. The
composite SED for BL Lac objects is adopted from \cite{fmc+98},
specifically from the sample characterized by radio luminosities of
$10^{42}$--$10^{43}$\,erg\,s$^{-1}$, which closely matches the SED of
\grs\ in the X-ray range; these data were recalculated to our adopted
value $H_0=75$\,km\,s$^{-1}$\,Mpc$^{-1}$.
}
\label{broad}
\end{figure}

We finally discuss the possibility that \grs\ is a BL Lac object,
as suggested by its positional coincidence with the gamma-ray
source \egret. There are two arguments against this
hypothesis. First, the optical spectrum of \grs\ is completely dominated by
broad lines characteristic of Seyfert~1s \citep{mmc+98}. This
contrasts with the usual situation for BL Lac objects, when it is very
difficult to discern any emission lines against the strong continuum
collimated toward us. Secondly, radio observations show only a weak
source at the \grs\ position, with a flux $\sim 10$\,mJy at 10\,GHz
\citep{mmc+98} . If \grs\ were a typical BL Lac object, one would
expect its radio counterpart to be some 3 orders of magnitude stronger
\citep{fmc+98}. This is demonstrated in Fig.~\ref{broad}, where
the broadband spectral energy distribution (SED) of \grs\ is compared
with the composite SED of BL Lac objects of
similar X-ray luminosity. On the other hand, the radio to X-ray
spectrum of \grs\ is very similar to that of the architypal Seyfert~1
galaxy NGC~4151 (see Fig.~\ref{broad}). 

Therefore, either 1) \grs\ is a quite unusual Seyfert~1 galaxy producing
strong gamma-ray emission or 2) it has no relation to \egret. Given
that the number density of hard X-ray sources 
detected by INTEGRAL in the central $5^\circ\times 5^\circ$ region of
the Galaxy is $\sim 0.5$ sq. deq$^{-1}$ \citep{rsv+04}, the
probability of finding by chance an INTEGRAL source within the 
error box of \egret\ is $\sim 50$\%. On the other hand, the
EGRET map \citep{hbb+99} of the central ($\sim 10^\circ\times 10^\circ$)
region of the Galaxy indicates that the probability of finding by chance an
EGRET source consistent with the position of \grs\ is only $\sim 3$\%. The
upcoming GLAST mission will be able to localize the gamma-ray source
down to $\sim 1^\prime$ and thus settle the issue of
\grs/\egret\ association. 

\bigskip
\noindent {\sl Acknowledgments.}
We thank Eugene Churazov for providing us the software for IBIS data
analysis. We acknowledge support from Minpromnauka 
(grant of President of Russian Federation NSH-2083.2003.2) and the
programme of the Russian Academy of Sciences ``Non-stable phenomena in
astronomy''. This research has made use of data obtained through the
INTEGRAL Science Data Center (Versoix), Russian
Science Data Center of INTEGRAL (Moscow), High Energy Astrophysics
Science Archive Research Center Online Service provided by the
NASA/Goddard Space Flight Center, and the NASA/IPAC Extragalactic
Database (NED) operated by the Jet Propulsion Laboratory,
Caltech. INTEGRAL is an ESA project funded by ESA member states
(especially the PI countries: Denmark, France, Germany, Italy,
Spain, Switzerland), Czech Republic and Poland, and with the
participation of Russia and the USA.

\end{document}